\documentclass[
    preprint,               
    superscriptaddress,    
    amsmath, amssymb,      
    aip,                   
    apl,                   
    floatfix               
]{revtex4-2}

\usepackage{graphicx}  
\makeatletter
\def\selectlanguage#1{}
\makeatother
\usepackage{dcolumn}       
\usepackage{bm}            
\usepackage{booktabs}
\usepackage[utf8]{inputenc} 
\usepackage{siunitx}       
\usepackage{tikz}
\usetikzlibrary{arrows.meta, decorations.pathmorphing}
\usepackage{xspace}
\usepackage[version=4]{mhchem}
\usepackage{hyperref}      
\usepackage{cleveref}
\crefname{figure}{fig.}{figs.} 
\Crefname{figure}{Figure}{Figures}
\usepackage{hypcap}
\newcommand{\SiN}{\texorpdfstring{\ce{Si3N4}}{Si3N4}\xspace}
\newcommand{\ER}{\texorpdfstring{\ce{Er^{3+}}}{Er3+}\xspace}
\newcommand{\sio}{\texorpdfstring{\ce{SiO2}}{SiO2}\xspace}

\hypersetup{
    colorlinks=true,
    linkcolor=blue,
    citecolor=blue,
    urlcolor=blue
}

\begin{document}

\title{Strong enhancement of \ER emission at room temperature in \SiN metasurfaces}

\author{Fengkai Wei}
\email{fengkai.wei@uni-jena.de}
\affiliation{Institut für Festkörperphysik, Friedrich-Schiller-Universität (FSU) Jena, 07745 Jena, Germany}

\author{Xinru Ji}
\email{xinru.ji@epfl.ch}
\affiliation{Institute of Physics, Swiss Federal Institute of Technology Lausanne (EPFL), CH-1015 Lausanne, Switzerland}

\author{Tobias J. Kippenberg}
\email{tobias.kippenberg@epfl.ch}
\affiliation{Institute of Physics, Swiss Federal Institute of Technology Lausanne (EPFL), CH-1015 Lausanne, Switzerland}

\author{Duk-Yong Choi}
\email{duk.choi@anu.edu.au}
\affiliation{Research School of Physics, The Australian National University, ACT 2601, Australia}

\author{Carsten Ronning$^{\ast}$}
\altaffiliation{Corresponding author}
\email{carsten.ronning@uni-jena.de}
\affiliation{Institut für Festkörperphysik, Friedrich-Schiller-Universität (FSU) Jena, 07745 Jena, Germany}

\date{\today}

\begin{abstract}
We report a significant enhancement of room-temperature photoluminescence from trivalent erbium-doped (\ER) silicon nitride (\SiN) metasurfaces at the telecommunication wavelength prepared via ion implantation. The metasurfaces, consisting of periodic nanocylinder arrays, are designed to support Mie-type resonances that tailor the local density of optical states. By systematically optimizing the nanocylinder radii, we achieve a photoluminescence (PL) enhancement factor of ${ }\approx 18$ at a radius of \SI{390}{\nano\metre} after thermal annealing, which is in excellent agreement with our simulations. Time-resolved PL measurements reveal a nearly ten-fold reduction in luminescence lifetime, confirming that the enhancement is primarily driven by the Purcell effect. Furthermore, we demonstrate that the PL intensity is strongly dependent on the \ER ion implantation depth, with a four-fold increase in emission observed from \SI{20}{\nano\metre} to \SI{80}{\nano\metre} ion range. These results provide a robust pathway for integrating efficient, active light sources into CMOS-compatible photonic device.
\end{abstract}


\maketitle 

\newpage

\section{Introduction}

Silicon nitride (\SiN) has emerged as a cornerstone material for modern photonic platforms due to its exceptional physical and optical properties, including low propagation loss, a wide transparency window extending from visible to mid-infrared, and a high refractive index that provides optical confinement\cite{compat2013NatPho}. Despite its structural and optical advantages, \SiN remains inherently a passive material lacking efficient light-emission capabilities. The lack of active, on-chip light sources limits the development of further functionalities\cite{OnchipLightSrc2015LightSciApp}. To the best of our knowledge, this work is the first report of room-temperature Er3+ emission enhancement in ion-implanted Si3N4 metasurfaces, with a systematic study of implantation depth and resonant LDOS engineering.

Rare earth (RE) element doping, one mainstream of introducing active functionalities into host materials, has been extensively explored\cite{IntEr3Emitters2025NanoMat}. Trivalent erbium ions (\ER) are of particular importance due to their intra-4f transitions at \SI{1.55}{\micro\metre}, which corresponds to the C-band telecommunication window\cite{Polman1997}. Despites its advantages, the room-temperature (RT) photoluminescence (PL) remains limited by low spontaneous emission and non-radiative paths in several host materials. Especially in silicon or many Si-base matrices, the erbium emission is only observed at low temperature, while at room-temperature condition, PL signals are significantly suppressed or even vanished\cite{Polman1997APL}. This dependence on low-temperature environments severely limits the application of erbium-incorporated devices in practical consumer-grade photonic chips and room-temperature communication systems.

To address these limitations, several nanophotonic architectures have been integrated with the Er:\SiN system to enhance light-matter interactions. Conventional strategies typically involve the use of long-path-length waveguides to increase the total interaction volume, or high-quality (Q) factor microcavities and photonic crystals to provide strong \ER emission\cite{LowEEr1992EleLet, ErWaveGuide2020OptExp, OptMicCav2003, CtrlDynEmis2004Nat}. Building upon these developments, dielectric metasurfaces have emerged as a versatile and compact platform for controlling emission through the precise patterning of subwavelength nanostructures\cite{OptResDie2016Sci, ResDieMeta2021APLMat}. By supporting optically induced electric and magnetic Mie resonances, these metasurfaces (MS) enable the engineering of the local density of optical states (LDOS) and the concentration of the electromagnetic field directly within active \ER-doped regions\cite{FuncMetaMie2017ACSPho}. The tailored electromagnetic environment facilitates a significant acceleration of the spontaneous emission rate via the Purcell effect\cite{Purcell1946}, offering a transformative route to overcome thermal quenching and achieve high-brightness luminescence even at room temperature\cite{PIC2022SCi, ErSiNPlas2009OptExp}.

In this work, we demonstrate a 18.1-fold room-temperature photoluminescence enhancement at \SI{1.53}{\micro\metre} by coupling ion-implanted \ER ions with Mie-resonant \SiN metasurfaces. This approach establishes a fully CMOS-compatible route to active nanophotonics, overcoming traditional thermal quenching limitations to achieve robust emission without cryogenic cooling or exotic host materials. By matching an optimal nanocylinder radius of \SI{390}{\nano\metre} with a simulated Purcell factor of 16.2, we prove that scalable ion implantation and simplified dielectric geometries can provide the high-performance light-matter interaction required for practical on-chip integration.

\section{Experimental}
\begin{figure}
    \centering
    \includegraphics[width=1\linewidth]{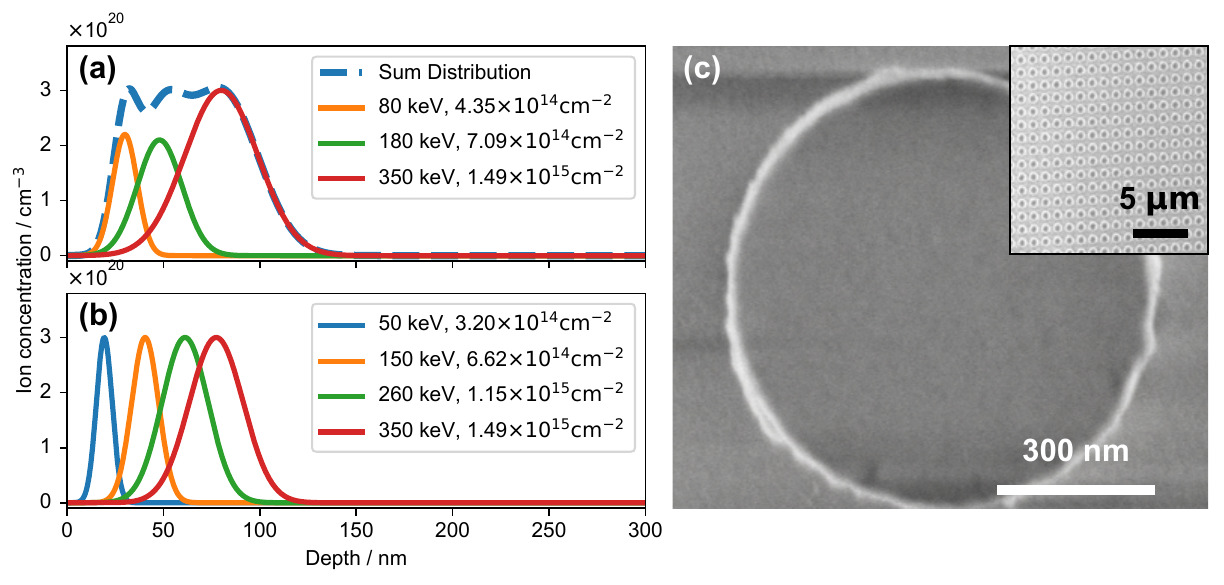}
    \caption{Ion implantaion profile of Er in \SiN for (a) 3-step implantation with vaying ion energies and (b) 4 different energies over depth simulated by Monte-Carlo SRIM code\cite{SRIM} simulations. (c) Scanning electron microscopy (SEM) images of fabricated metasurfaces (inset: overview of metasurface)}
    \label{fig:1}
\end{figure}

Silicon nitride (\SiN) thin films with thicknesses of \SI{365}{\nano\metre} were deposited on fused silica (\sio) using low-pressure chemical vapor deposition (LPCVD)\cite{LPCVDSiN2024Optica}. The resulting platform features a low surface roughness ($\approx \SI{0.3}{nm}$ root mean square for as-deposited films), stable refractive index ($n = \text{1.986}$ at \SI{1550}{\nano\metre}), and high thermal stability. In the numerical modeling, a slightly higher effective refractive index ($n=2.01$) is used for the Er-implanted active region to account for the changes in material composition, local density, and polarizability induced by ion implantation and rare-earth incorporation. The nitride hydrogen bond (N-H) and silicon hydrogen (Si-H) bonds eliminated by high temperature annealing at \SI{1200}{\degree}. At room temperature, erbium was implanted into the \SiN layer. A three-step implantation process was employed to achieve a uniform doping concentration of approximately \SI{3}{\times10^{20}cm^{-3}} across a depth range of 20nm to 100nm (see \cref{fig:1} (a)). For depth-dependency studies, individual samples were implanted at specific ion energies\cite{LowEEr1992EleLet} to precisely control the vertical distribution of the implanted erbium (see \cref{fig:1} (b)). The implantation profiles were simulated using the Monte-Carlo code "Stopping Range of Ions in Matter (SRIM) simulations"\cite{SRIM}, and the respective ion fluences are listed in the insets of \cref{fig:1}.
To recover the implantation-induced damage and optically activate the erbium ions, the fabricated devices were subjected to thermal annealing in a nitrogen atmosphere at \SI{500}{\degree C} for 1 hour.

The metasurface consists of a periodic array of \SiN nanocylinders designed to support Mie-type resonances\cite{KerNanoMeta2018OptExp}. The cylinders with a fixed height of \SI{365}{\nano\metre} were arranged in a square lattice with a period of \SI{1050}{\nano\metre} and the radius was varied between \SI{240}{\nano\metre} and \SI{465}{\nano\metre} with a stepsize of \SI{15}{\nano\metre}. The patterns were defined using electron beam lithography (EBL). A \SI{500}{\nano\metre} thick layer of positive resist (ZEP520A) was spin-coated and baked at \SI{180}{\degree C} following EBL exposure at an acceleration voltage of \SI{125}{kV} (ELS-Boden 125), a \SI{50}{nm}-thick aluminum (Al) film was deposited via E-beam evaporation(Temescal BJD-2000) . The patterns were then transferred to the \SiN layer through inductively coupled plasma reactive ion etching (ICP-RIE) using a \ce{CHF3}/\ce{O2} plasma chemistry. After \SiN etching, the remaining Al mask was removed by an aluminum etchant type A (Sigma-Aldrich) at \SI{80}{\degree C}, followed by cleaning with deionized water and isopropanol. \Cref{fig:1} (c) shows an scanning electron microscopy (SEM) image of the fabricated metasurface, confirming the high fidelity of the pattern transfer and the uniformity of the nanocylinder arrays.

Theoretical modeling of the metasurface was carried out using the finite-difference time-domain (FDTD) method\cite{Yee1966} via the Ansys Lumerical FDTD software, version 2024 R1\cite{LumericalFDTD}, and finite element method (FEM)\cite{Jin2014} implemented in COMSOL Multiphysics\textregistered{}, version 6.2\cite{COMSOL}. \ER ions were modeled as dipoles at \SI{1550}{\nano\metre} placed at various orientations and positions within the \SiN ($n=2.01$ at \SI{1550}{\nano\metre}) nanocylinders on bulk \sio ($n=1.45$) to account for the spatial distribution of the emitters. Periodic boundary conditions (PBC) and perfectly matched layers (PML) were implemented to simulate an infinite array and absorb outgoing radiation, respectively. The Purcell factor was calculated by normalizing the total power radiated by a dipole in the metasurface environment to its radiation in a bulk \SiN medium.

Room-temperature PL measurements were performed using a custom-built micro-PL setup. A continuous-wave (CW) laser at \SI{520}{\nano\metre} served as the excitation source with adjustable power ranging from \SI{1}{mW} to \SI{80}{mW} focused onto the sample via a 20$\times$ objective ($\text{NA}=0.65$) yielding a spot size of approximately \SI{6}{\micro\metre}. The backward-emitted PL was collected by the same objective, filtered through a \SI{1060}{\nano\metre} dichroic mirror and a \SI{1400}{\nano\metre} long-pass filter to reject the pump signal, and finally analyzed by a NIRQuest+1.7 spectrometer. Power-dependent PL was measured by varying the excitation power from \SI{5}{mW} to \SI{80}{mW} to investigate the emission saturation behavior. The \ER lifetime at \SI{1530}{\nano\metre} was measured using a pulsed laser source and a APD410C/M InGaAs avalanche photodetector. The power of the laser source was modulated at a frequency of \SI{250}{Hz}. 

\section{Results and discussion}
\begin{figure}
    \centering
    \includegraphics[width=1\linewidth]{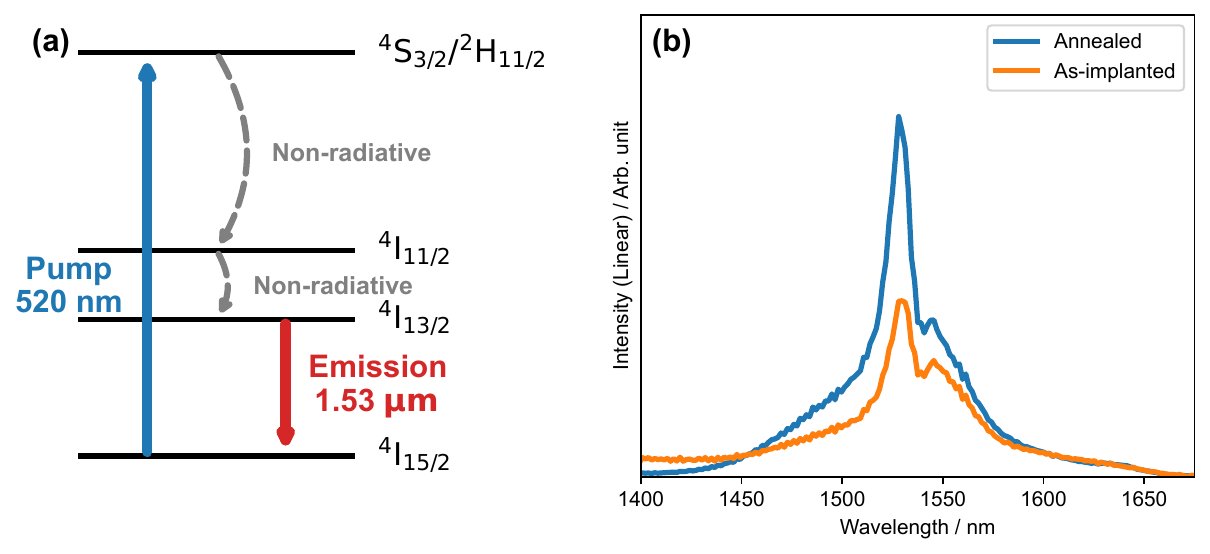}
    \caption{(a) Energy level diagram of \ER with excitation and relaxation paths; (b) PL of as-imlpanted (orange) and annealed (blue) planar thin film \SiN excited by \SI{520}{nm} at \SI{50}{mW}.}
    \label{fig:2}
\end{figure}

Before investigating the metasurface response, we first characterized the PL properties of the planar Er-implanted \SiN films at RT to establish a baseline. The emission of \ER at C-band involves a multi-step process\cite{StrVisInfra2008JAP}: ions are first excited from the ground state ($^4\text{I}_{15/2}$) to the higher-energy $^4\text{S}_{3/2} / ^2\text{H}_{11/2}$ manifolds by the \SI{520}{\nano\metre} pump, followed by rapid non-radiative relaxations to the $^4\text{I}_{13/2}$ metastable state, from which the radiative transition occurs (see \cref{fig:2} (a)).

A key focus of this study was the optimization of RT emission. We measureed the RT-PL intensity of the as-implanted planar thin film, and \cref{fig:2} (b) shows the resulting spectrum (orange). The spectrum features two peaks and broadening. A main emission peak at \SI{1530}{\nano\metre}, which is well known as the $^4\text{I}_{13/2}$ to $^4\text{I}_{15/2}$ transition of \ER ions. A side peak at \SI{1550}{\nano\metre} results from the Stark splitting of the energy levels due to the local crystal fields in the \SiN matrix. The overall broadening from \SI{1400}{\nano\metre} to \SI{1670}{\nano\metre} comes from the amorphous nature of \SiN, due to lack of long-range order, each \ER ion has slightly different local bond angle, length and coorination number. The emission is a superposition of all \ER ions with sharp peaks with deviations, which renders a smooth, broadened envolope in spectrum. Similar spectra were also observed in Er-doped silicon-rich nitride films\cite{ERinSRN2009OptExp} and Er-doped SiO thin films\cite{ERinSiO2007JAP}. In the as-implanted state, the \SiN matrix contains "defects" and non-radiative recombination centers induced by the high-energy ion bombardment. Here, dangling bonds, which typically forms in the amorphous \SiN matrix upon ion irradiation create sub-states in the band gap, which strongly compete with the \ER radiative transition. Thus, a thermal treatment is essential for activating further \ER ions and mitigating the non-radiative recombination centers\cite{Polman1997}. Consequently, the PL intensity of the planar samples exhibited a 2-fold increase compared at \SI{1530}{nm} to its as-implanted state after annealing to \SI{500}{\degree C}, as shown in \cref{fig:2} (b).

\begin{figure}
    \centering
    \includegraphics[width=1\linewidth]{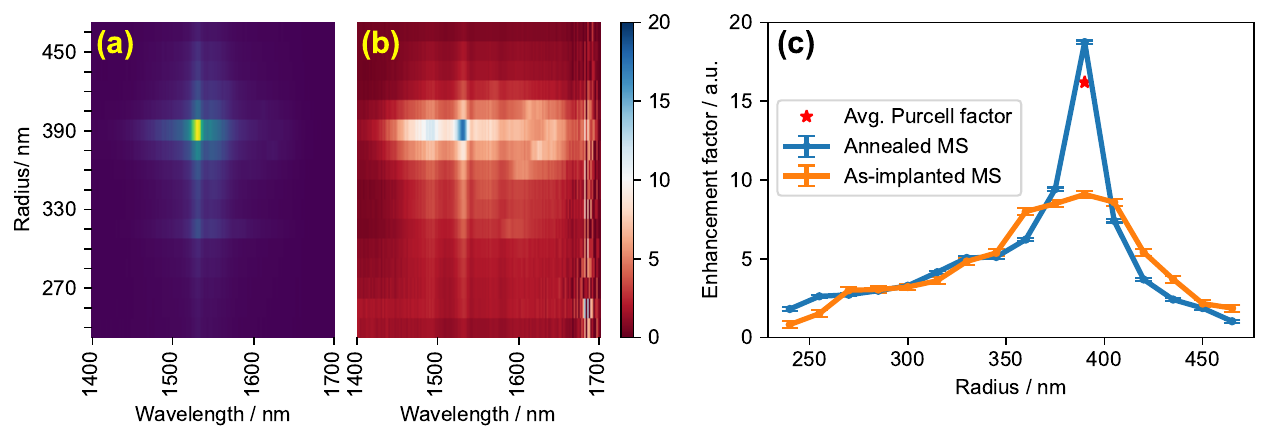}
    \caption{Heatmaps of (a) PL spectra (normalized to maximum intensity) and (b) relative enhancement factor by annealed MS as a function of radius; (c) Radius dependence of relative enhancement factor at wavelength = \SI{1530}{nm}}
    \label{fig:3}
\end{figure}

Next, the PL spectra of fabricated \ER doped \SiN metasurfaces consisting of a periodic array of nanocylinders were measured as a function of radius. The spectra of annealed samples were stacked vertically and are shown as a heatmap in \cref{fig:3} (a). For the determination of the relative enhancement factor, the spectra were normalized to the spectra of the planar thin films and corrected with the area fraction that was removed by the EBL and etching around the \SiN nanocylinders. The corresponding heatmaps are shown in \cref{fig:3} (b), and \cref{fig:3} (c) shows the emission enhancement at a wavelength of \SI{1530}{\nano\metre} of the \ER implanted \SiN metasurface as a function of the radius. As shown in \cref{fig:3}, the room temperature PL enhancement exhibit a clear dependence on the nanocylinder radius. Peak enhancement of PL is achieved at a radius of $R ={ }$\SI{390}{\nano\metre} for both as-implanted and annealed samples. The as-implanted MS yields an enhancement of 8.8 and the annealed MS even drastically boosted the value to 18.1. Beyond the magnitude of enhancement, the PL response of annealed samples shows a pronounced and narrower resonance peak compared to the as-implanted counterparts. This behavior can be attributed to the dual effect of thermal annealing on both the host matrix and the emitters. Firstly, the thermal treatment effectively repairs implantation-induced damage and passivates non-radiative recombination centers, such as silicon dangling bonds, which significantly enhances the internal quantum efficiency of the \ER ions. Secondly, the mitigation of these defects reduces parasitic optical absorption within the \SiN nanostructures, thereby increasing the quality factor (Q) of the Mie-type resonances. In the as-implanted sample, implantation-induced damage and non-radiative centers introduce additional loss and spectral broadening, which washes out the resonance and spreads the enhancement over a wider radius range. After annealing, this extra loss is suppressed, the Mie resonance becomes spectrally sharper, and the electromagnetic field localization inside the nanocylinder is strengthened. Consequently, a higher Q-factor leads to a more spatially and spectrally sensitive coupling between the emitters and the electromagnetic modes, manifesting as the observed narrowing of the resonance peak in the geometric parameter space. This also explains the relatively small difference between annealed and as-implanted metasurfaces away from $R={ }$\SI{390}{\nano\metre}. In those off-resonant radii, the PL is primarily limited by poor mode overlap rather than by the internal quantum efficiency of the emitters, so the benefit of annealing is less visible. In contrast, the planar sample in \cref{fig:2}(b) is not resonantly enhanced, and its signal therefore directly reflects the improved emitter activation and reduced non-radiative loss after annealing. At $R={ }$\SI{390}{\nano\metre}, both the emitter efficiency and the metasurface resonance are optimized, so the annealed sample shows a substantially larger enhancement compared to the as-implanted case. This is why the annealed MS shows a very steep enhancement peak at $R={ }$\SI{390}{\nano\metre}: the improved resonance quality makes the Purcell enhancement much more sensitive to small radius variations, so only the structures closest to the optimal resonant condition exhibit the full 18.1-fold boost. In other words, the peak at 390 nm corresponds to the most favorable overlap of the resonant mode, ion depth distribution, and emission wavelength; a small deviation in radius detunes the Mie modes sufficiently that the local density of states drops rapidly, producing the sharp peak observed in \cref{fig:3}(c).
\begin{figure}[ht]
    \centering
    \includegraphics[width=1\linewidth]{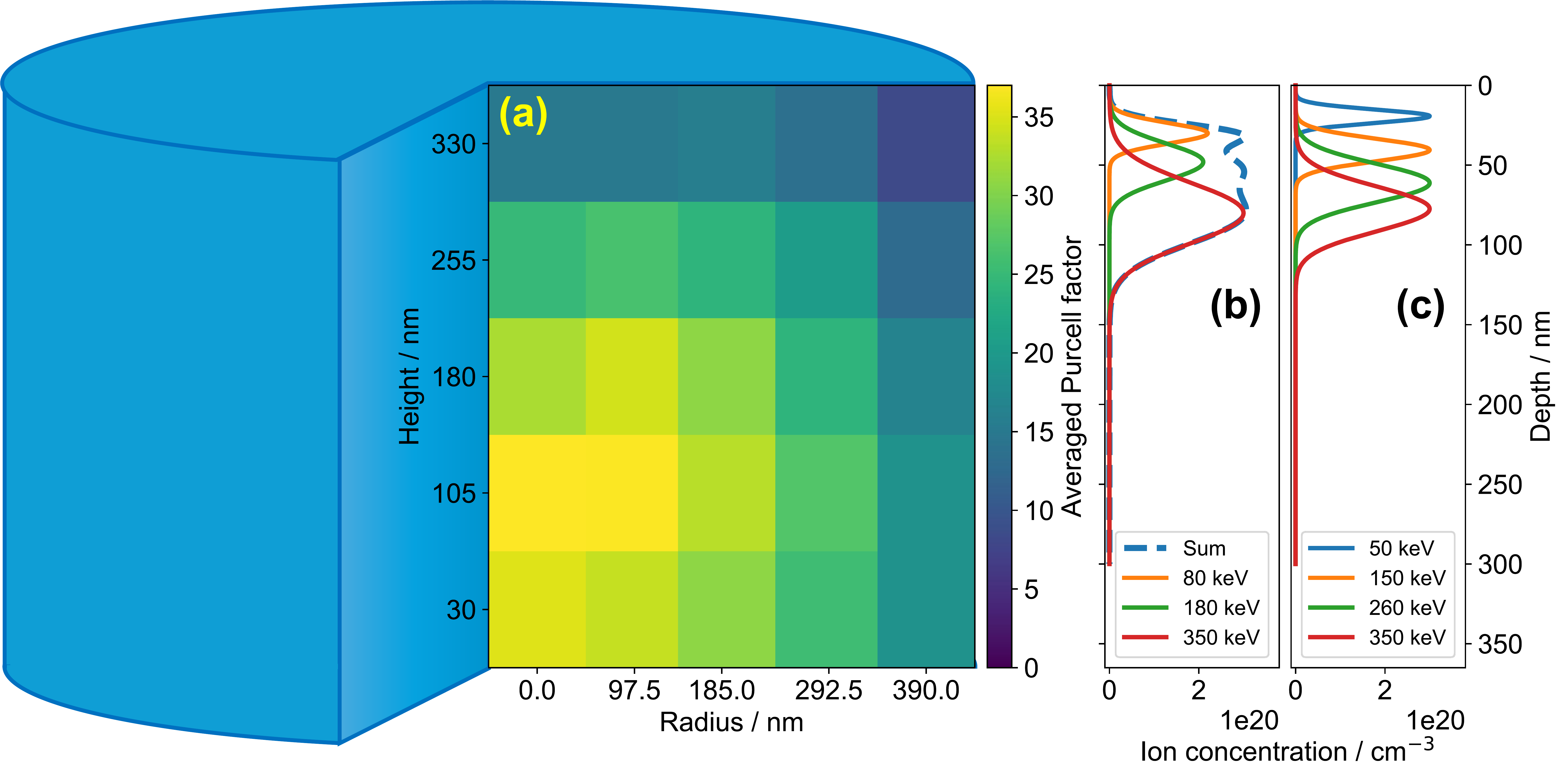}
    \caption{(a) FDTD simulated Purcell factor as a function of site. SRIM simulated \ER depth of the (b) 3-step implantation and (c) 4 different ion energies.}
    \label{fig:4}
\end{figure}

Notably, the experimental measurement results show an excellent agreement with our numerical simulations, which are displayed in \cref{fig:4}. The spatial distribution of the Purcell factor in the nanocylinder is depicted in \cref{fig:4} (a). The cross-sectional heatmap highlights a prominent core-loading characteristic: the maximum Purcell enhancement is achieved at the axial center ($R ={ }$\SI{0}{\nano\metre}) and a height of $h ={ }$\SI{105}{\nano\metre}. Moving away from this optimal position in either the radial or axial direction results in a gradual decline of the Purcell factor. The simulated Purcell factor averages the contributions from both electric and magnetic dipole orientations across the implantation profile. After averaging all distributed regions of \ER simulated by SRIM (see \cref{fig:4} (b)) an averaged value of the Purcell factor obtained approximately 16 at a radius of \SI{390}{\nano\metre}, which is also shown as a red asterisk in \cref{fig:3} (c). This close correspondence between the theoretical prediction and the measured 18.1-fold enhancement validates that the observed performance is driven by the excitation of discrete Mie-resonance modes within the nanocylinders. The slight discrepancy between experimental (18.1) and simulated (16.2) values may arise from the simplified idealized model with only 3 orientations of emitting dipole, comparing with random case of dioples with all orientations.

To further elucidate the physical mechanism underlying the observed PL enhancement and confirm the contribution of the Purcell effect, room-temperature time-resolved PL (TRPL) measurements \cite{PrinNano2012Book} were performed with varying radii. 
\begin{figure}[ht]
    \centering
    \includegraphics[width=1\linewidth]{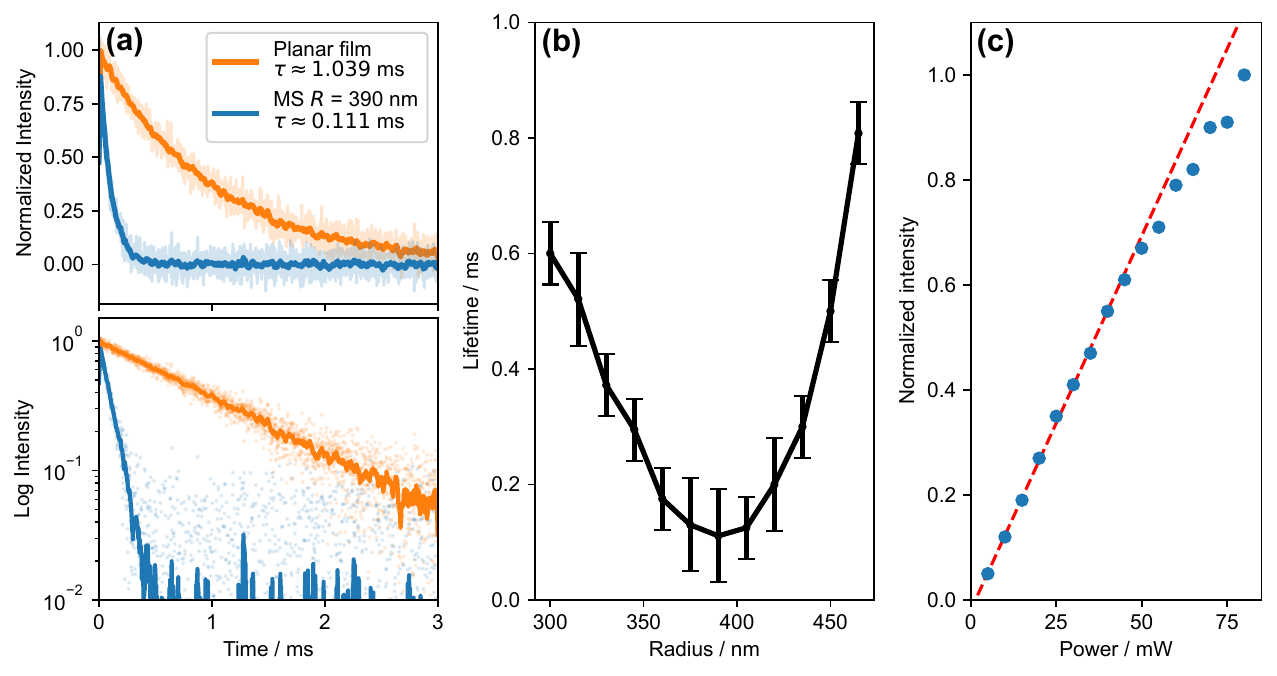}
    \caption{(a) \ER lifetime measurement of an implanted planar thin film (blue) and MS (orange) with radius $R ={ }$\SI{390}{\nano\metre}, top and bottom subplots are linear and logarithmic scale, respectively. Light and dark lines are denoised and original results repectively. (b) \ER lifetime in MS as a function of radius. (c) Normalized PL intensity as a function of excitation power of annealed MS with $R ={ }$\SI{390}{\nano\metre}. Red dashed line is linear regression of data points of low power values.}
    \label{fig:5}
\end{figure}
As illustrated in \cref{fig:5} (a), a direct comparison between the \ER ions implanted in the MS with radius $R ={ }$\SI{390}{\nano\metre} and those in a planar thin film reveals a dramatic acceleration in the decay time. While the planar control region exhibits a characteristic lifetime of $\tau\approx{ }$\SI{1.039}{ms}, a typical lifetime of \ER in \SiN matrix\cite{ErLifetimeSiN2009APL}, the metasurface with a radius of $R ={ }$\SI{390}{\nano\metre} reduces this value to $\approx$\SI{0.111}{ms}. This nearly ten-fold reduction in lifetime is a clear signature of the Purcell effect, where the LDOS is significantly enhanced by the resonant modes of the metasurface. The log-scale representation in the lower panel of \cref{fig:5} (a) further confirms that the decay is a single expoential decay, indicating a significant fraction of emitters is effectively coupled with the metasurface resonance mode.

The dependence of the \ER lifetime on the radius from $R ={ }$\SI{300}{\nano\metre} to $R ={ }$\SI{465}{\nano\metre} of the metasurface is further plotted in \cref{fig:5} (b). Starting from 300 nm, the lifetime decreases steadily as the radius increases, reaching a minimum of approximately $\tau \approx{ }$\SI{0.111}{ms}, at $R ={ }$390 nm. This minimum represents the optimal "on-resonance" condition, where the metasurface resonant modes couple most effectively with the \ER emission transition. As the radius deviates from \SI{390}{\nano\metre}, the lifetime begins to recover, which is attributed to the spectral detuning of the cavity mode away from the \ER transition. 

The excitation dynamics and the efficiency of the PL process were evaluated by examining the power dependence of the emission intensity, as shown in \cref{fig:5} (c). 
In the low-excitation regime, the PL intensity exhibits an almost linear correlation with the excitation power, as indicated by the dashed red line. This linearity implies that the system operates far from saturation, where the emission rate is directly proportional to the excitation rate. However, as the pump power exceeds \SI{40}{mW}, the experimental data points begin to deviate slightly from the linear fit, showing a clear sub-linear progression. This behavior is indicative of the onset of power saturation, occurring when the pump rate becomes comparable to the total relaxation rate of the \ER ions. At these elevated power levels, the depletion of the ground-state population limits further increases in PL intensity. Despite the onset of saturation at higher powers, the MS maintains a robust emission profile, demonstrating its effectiveness in concentrating the electromagnetic field to achieve high-brightness luminescence even under moderate pump intensities.

\begin{figure}
    \centering
    \includegraphics[width=1\linewidth]{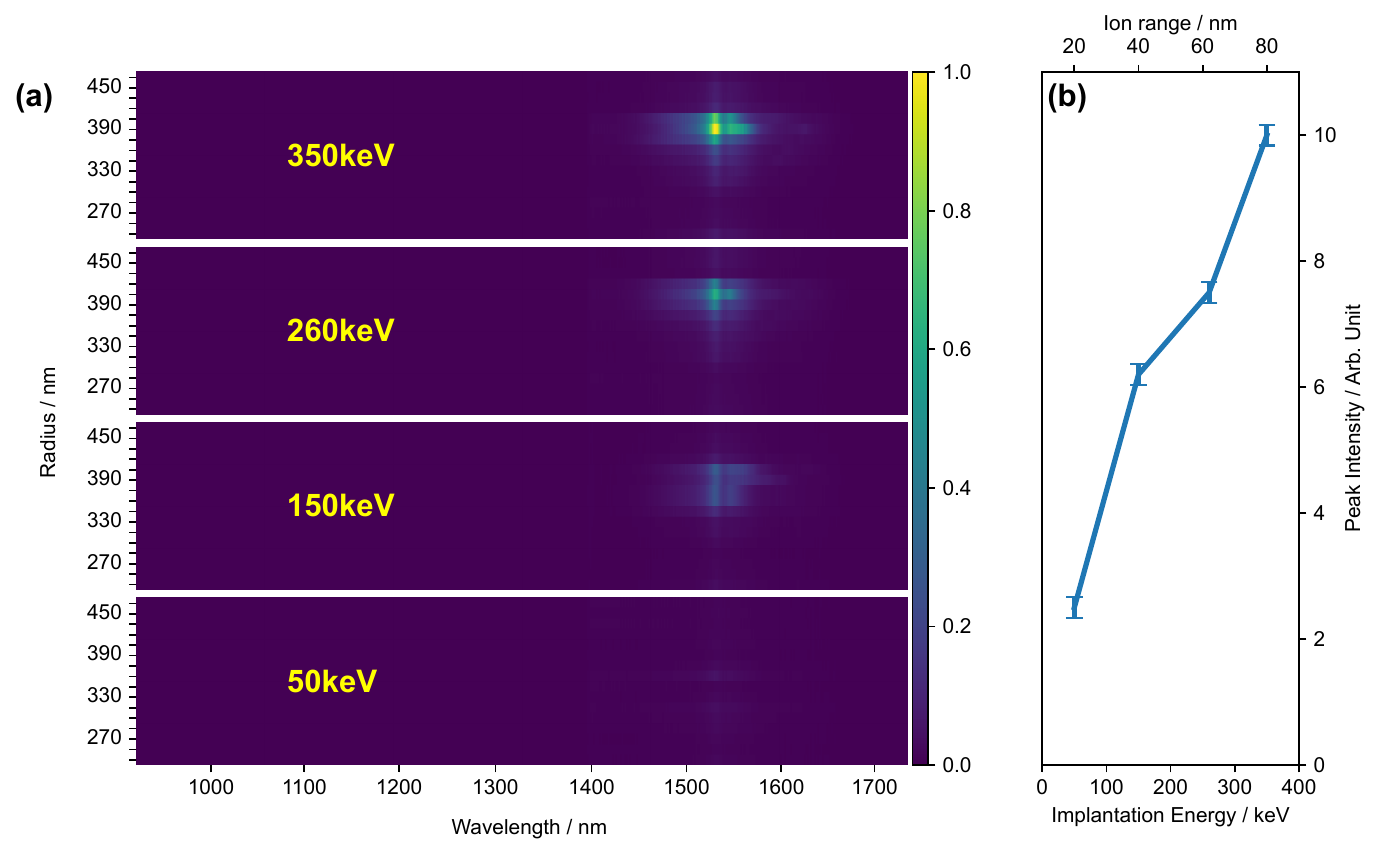}
    \caption{PL of depth-dependent \ER doped metasurfaces. (a) PL spectral heatmaps for varied nanocylinder radii at implantation energies from \SI{50}{keV} to \SI{350}{keV}. (b) Peak PL intensity at the resonant radius (R=\SI{390}{\nano\metre}) as a function of implantation energy.}
    \label{fig:6}
\end{figure}

A depth dependent distribution of the \ER emitters in the MS was realized by signgle ion implantations with varying ion energy. \Cref{fig:6} (a) displays PL spectral heatmaps for different metasurface radii as a function of implantation energy 50 keV, 150 keV, 260 keV and 350 keV, respectively.
Across all energies, a clear PL enhancement is observed around \SI{1530}{\nano\metre} for radii centered at approximately \SI{390}{\nano\metre}, aligning with the previously shown metasurface resonance condition. Furthermore, the brightness of this peak is highly dependent on the ion energy (ion range), with significantly stronger PL intensity observed at higher implantation energies (\SI{350}{keV} and \SI{260}{keV}) compared to the lowest energy (\SI{50}{keV}).
This trend is quantified in \cref{fig:6} (b), which plots the peak PL intensity at MS radius $R ={ }$\SI{390}{\nano\metre} as a function of the implantation energy and corresponding ion range (implantation depth). The peak intensity increases monotonically to a 4-fold enhancement maximized at \SI{350}{keV}. 
To correlate experimental performance with simulation, the SRIM simulations of implantation depth profiles for various energies were  illustrated in \cref{fig:4} (c). As the implantation energy increases from \SI{50}{keV} to \SI{350}{keV}, the peak ion concentration shifts deeper into the substrate. By comparing the ion distributions with the Purcell factor mapping in \cref{fig:4} (c), it becomes evident that higher implantation energies effectively "fill" the high-Purcell-factor region. While the 80 keV ions are confined to the top \SI{40}{\nano\metre}, they occupy a smaller total volume. In contrast, the \SI{350}{keV} implantation results in a significantly higher integrated erbium ions across the optimal 0-\SI{150}{\nano\metre} height range. The higher PL intensity at \SI{350}{keV} is thus not merely a result of increased erbium ion concentration, but rather the better positioning of a high-density the erbium ensemble within the high Purcell factor zone. Consequently, the depth-dependent study confirms that achieving maximum device brightness requires a precise balance: ions must be implanted deep enough to couple with metasurface resonant modes\cite{CtrlDynEmis2004Nat}.

\section{Conclusion}
In summary, we have successfully demonstrated enhancement of the room temperature \SI{1530}{\nano\metre} emission from \ER doped \SiN metasurfaces. Our findings show that the excitation of Mie-type resonances in \SiN nanocylinders can significantly concentrate electromagnetic fields and modify the spontaneous emission rate of embedded \ER ions. The optimal metasurface design yields an 18.1-fold PL intensity boost, supported by both steady-state spectroscopy and time-resolved lifetime measurements, which validates the dominance of the Purcell effect in this architecture. Notably, we revealed that the PL enhancement is highly sensitive to the vertical positioning of emitters, where deeper implantation depths leads to superior spatial overlap with the resonant modes compared to shallow doping. The demonstrated compatibility with standard lithographic processes and the effectiveness of thermal annealing for defect recovery highlight the potential of this platform. For future work, we aim to position the erbium ions deeper within the \SiN nanocylinders; this will be pursued either by using higher acceleration voltages or by depositing a thin layer to create a layer-implant-layer geometry. This work facilitates the development of high-brightness, on-chip light sources, bridging the gap between passive \SiN photonics and active telecommunication-grade optoelectronic devices.

\section{Acknowledgement}
This work was supported by the International Research Training Group (IRTG 2675) "Meta-Active" Project 437527638, funded by the German Research Foundation (DFG). The authors would like to acknowledge the technical assistance and access to fabrication facilities provided by the Australian National Fabrication Facility (ANFF). Furthermore, this research received funding from the European Research Council within the ERC Synergy project ATHENS (101167540).
\bibliographystyle{aipnum4-2} 
\bibliography{ref_26032026}

\end{document}